\documentclass[11pt]{llncs}

\newcommand{\ignore}[1]{}

\usepackage{graphicx}

\vfuzz \hfuzz                                          
\newcounter{step} 
\newcommand{\nostep}{
\item[] \addvspace{5pt}}
\newcommand{\step}[1][]{
\item[] \addvspace{5pt}
\refstepcounter{step}
{\bf \arabic{step}#1.}\ }
\newcommand{\debprotocol}[1]{
\begin{center}{\bf #1}
\end{center}
\begin{list}{}{
\setlength{\leftmargin}{0pt}
}}

\newcommand{\finphase}{\end{list} \mbox{}}
\newcommand{\finprotocol}{\end{list} \setcounter{step}{0}}

\title{General Security Definition and Composability\\ for Quantum \&
Classical Protocols}

\titlerunning{Composability for Quantum \& Classical Protocols} 
\author{ Michael Ben-Or \inst{1} \and Dominic
Mayers\inst{2}
}
\institute{ Hebrew University, Jerusalem, Israel ,\,
\email{benor@cs.huji.ac.il} \and IQI, California Institute of
Technology ,\, \email{dmayers@cs.caltech.edu} }

\authorrunning{M.~Ben-Or \and D.~Mayers}

\begin{document}

\maketitle
\mbox{}
\vspace{-.3in}

\begin{abstract} We generalize the universally composable
definition of Canetti to the Quantum World. The basic idea is the same
as in the classical world.  The main contribution is that we unfold
the result in a new model which is well adapted to quantum protocols.
We also simplify some aspects of the classical case. In particular,
the case of protocols with an arbitrary number of layers of
sub-protocols is naturally covered in the proposed model.
\end{abstract}
\vspace{-.2in}
\pagestyle{plain}

\section{Introduction}\label{intro}
In recent times, the analysis of complex cryptographic protocols has
been getting more and more attention. A complex cryptographic protocol
is a protocol that has a tree of sub-protocols in which any
sub-protocol in the tree calls its children. The leafs are the
primitives.  In all approaches for the analysis of complex
cryptographic protocols, one simply obtains security results for the
primitives, then use these security results to obtain security results
for the parent sub-protocols, and so on until one reaches the root of
the tree. In all approaches, this might be done over a long period of
time by many researchers that each publishes their respective results.
In all approaches, a result about a given sub-protocol can be used in
a large class of (complex) protocols that call this sub-protocol. The
different approaches only differ in the type of security results that
are obtained for these individual sub-protocols.

Essentially, Canetti's universally composable definition
\cite{Canetti01} states that we can replace the sub-protocol with an
associated ideal protocol together with a simulator, and the
environment of the protocol (which includes the parent sub-protocol in
the tree) will not notice the difference. The key point that explains
the power of this approach is that, whenever we consider a
sub-protocol in the tree, the lower level sub-protocols in the tree
are already replaced by their associated ideal protocols. So, at every
stage of the proof, we only consider sub-protocols that calls ideal
protocols, and this makes a big difference when we try to prove the
security of a protocol. Note that Canetti's model only covered a tree
of constant dept because of a technical issue that we will explain
later.

We unfold this basic principle in a different model that is more
adapted to quantum protocols.  The idea that the universally
composable theorem of Canetti is also valid in the quantum world was
previously suggested in \cite{CrepeauGottesmanSmith02}.  However, no
proof of the theorem or even a model to state the theorem in the
quantum world was provided. The quantum scenario has lead to many
surprises in the past, especially because of the extra difficulty
associated with entanglement, and so it was necessary to unfold
carefully the universal composability theorem and the associated model
in the quantum world. Our work was first reported in
\cite{BenOrMayers02}.  A protocol in our model is a set of circuits
executed by distinct participants. A circuit is a set quantum
registers and a partially ordered set (POSET) of quantum gates on
these registers.  As we will see in more details later (see section
\ref{BasicModel}), the POSET can depend on the value of control
registers. As in the classical case, the model must also define
environments, ideal protocols and simulators.

Just deciding what is a useful security result for a sub-protocol in
the tree (really only how to define the result, not how to obtain it)
is already a difficult challenge.  The principle of simulatability was
proposed in the early 80's to address this challenge
\cite{Yao82}. Essentially, this principle says that if the view of the
corrupted parties can be efficiently simulated by a machine which
interacts with an ideal protocol, then these corrupted parties did not
defeat the protocol.  So, the idea that the analyzed protocol must be
equivalent in someway to an ideal protocol was already there in the
simulatability principle. This principle was used in the early 80's to
define and prove the security of multiparty computation protocols
\cite{Yao82}, but no composability result was proven at the time.  The
simulatability principle was successfully used again later
\cite{GolswasserLevin90,Beaver91,MicaliRogaway91} and issues such as
composition were considered and partial composition theorems were
obtained \cite{Beaver91,MicaliRogaway91}.

The universally composable definition of Canetti and its associated
universal composability theorem only came a few years ago
\cite{Canetti01}. Independently, people at Zurich IBM obtained
composability results in a framework that interestingly can be used
with formal methods to automate security proofs
\cite{PfitzmannWaidnerA00,PfitzmannWaidnerB00}.  Impressively, with
independent work and under the inspiration of \cite{Canetti01}, they
recently obtained a universal composability theorem
\cite{BackesPfitzmannWaidner04} in this framework.  At the least at
this stage, we use the more informal framework of Canetti. The key
element is that these definitions were proven useful (for examples see
\cite{CanettiFishlin01,CanettiKrawczyk02,CanettiLindellOstrovskySahai02,PfitzmannWaidnerB00,Wikstrom04,DamgardFehrMorozovSalvail04}).

The notion of security in the quantum world has its own history which
we briefly review in Appendix \ref{history}.  Essentially, the notion
of security in the quantum world over the last 20 years used an
unstructured or unrestricted framework for security proofs.  In this
unrestricted framework, there is no rule except common sense and
mathematics: a security definition can be any property of an analyzed
protocol that is potentially useful (when we traverse the tree
bottom-up as previously mentioned).  The purpose of this work, first
reported in \cite{BenOrMayers02}, is to introduce the more structured
framework of universal composability in the quantum world.  We will
give an example of the difficulties that arise in the unrestricted
framework in the next paragraph, but it will never replace the
understanding that many researchers gained through the experience of
proving or trying to prove the security of complex protocols (for
example see
\cite{CanettiFishlin01,CanettiKrawczyk02,CanettiLindellOstrovskySahai02,PfitzmannWaidnerB00,Wikstrom04,DamgardFehrMorozovSalvail04}.)
The point is that the universal composability framework is not just a
general proposal for ``nice'' security definitions. Whether a security
definition is nice or not nice is a matter of taste. The universal
composability framework is a practical framework that allows security
results that seem otherwise difficult to achieve. Moreover, it is less
error prone.

Of course, our point is that this general framework is also very much
needed in the quantum world, perhaps even more than in the classical
word in view of the additional difficulties related to quantum
entanglement. As an example, consider the following key degradation
problem. Quantum key distribution protocols requires classical
authenticated channels. All security proofs for quantum key
distribution assume that ideal classical authenticated channels are
used. However, in practice, real classical authentication protocols
are used, because there is no such a thing as an ideal authenticated
classical channel. The difficulty is that classical authentication
protocols require a small private key initially and very often this
small private key will come from a previous quantum key distribution
protocol. Let us assume that sometimes in the past, Alice and Bob met
to exchange an ideal small private key and then went into an
alternating sequence of authentication and quantum key distribution
protocols, each protocol in this sequence calling the previous one. In
this case the tree of sub-protocols has only one branch which
corresponds to this sequence. The question is whether the key
generated by this complex quantum key distribution protocol, this one
branch tree, is still secure after $n$ calls to classical
authentication and quantum key distribution.  This is just an
example. The fact is that quantum key distribution will be used in
many other types of application protocols. The universal composability
of quantum key distribution is analyzed in
\cite{BenOrHorodeckiLeungMayersOppenheim04}. Obviously, a similar
question can be raised for other quantum protocols than quantum key
distribution.

To summarize, our contribution is to unfold the framework of universal
composability in a model that is well adapted to quantum protocols.
It is also an interesting alternative model for the universal
composability of classical protocols as well.

\section{Basic concepts} \label{basic_concepts}
In this section, we discuss examples of the five main concepts, the
concepts of protocol, application, adversary, ideal protocol and
simulator, and we give the ideas that are crucial in the composability
theorem.

\subsection{An example of protocol}
We start with an example that is all classical, but it will be enough
to explain the basic idea. It is a bit commitment from Alice to Bob
that calls an ideal ${2 \choose 1}$-String-OT from Bob to Alice. In a
${2 \choose 1}$-String-OT protocol from Bob to Alice, Bob has two
strings $s[1]$ and $s[2]$ as inputs whereas Alice as a bit $c$ as
input. The objective is that Alice receives the string $s[c]$, but
knows nothing about the other string whereas Bob does not know the bit
$c$, but gets an acknowledgement that Alice has chosen the bit $c$.
An ideal ${2 \choose 1}$-String-OT protocol from Bob to Alice can be
constructed with the help of a trusted party, which we call Charlie.
Bob sends the two strings to Charlie. Charlie receives the bit $c$
from Alice and sends the string $s[c]$ to Alice. Charlie also
acknowledges to Bob that Alice has chosen $c$, but he keeps $c$
private. This ${2 \choose 1}$-String-OT protocol is formally defined
in Appendix~\ref{formal_protocols}.

In the bit commitment protocol, Bob and Alice call the
${2 \choose 1}$-String-OT protocol.  To commit a bit $x$ Alice chooses
the string $\hat s = s[x]$ and Bob notes the acknowledgement. The
acknowledgement (which is output by Bob in the string-OT protocol) is
an important part of the commitment phase because the protocol would
not be binding otherwise: Alice could wait until much after the end of
the commitment phase to choose the bit $x$ in the
${2 \choose 1}$-String-OT protocol.  Later, to open the bit $x$, Alice
announces $x$ and the string $\hat s$. Bob checks that $\hat s = s[x]$
to accept or reject the bit.  This bit commitment protocol without the
${2 \choose 1}$-String-OT protocol is called a module. This module
calls the ${2 \choose 1}$-String-OT protocol, but it does not contain
it.  This module together with the ${2 \choose 1}$-String-OT protocol
is the (complete) bit commitment protocol.  The bit commitment module
is formally defined in Appendix~\ref{formal_protocols}.

Thus far, we used the names Alice, Bob and Charlie to refer to some
circuits (also called role-circuits) that are executed inside a
protocol.  However, a participant executes role-circuits in more than one
protocol. For example, Alice executes a circuit in the bit commitment
protocol that we call Alice, but different protocols have different
circuits with the name Alice.  To avoid any confusion, in the future,
we will often use the notation {\tt SOT-Alice}, {\tt BC-Alice}, etc.\
to distinguish these different circuits.  This becomes important when
we have a complex protocol with many sub-protocols (i.e. many
modules), and this is typically the situation where composable
security definitions are very useful.

\subsection{Application protocols}
A composable security definition essentially states that the analyzed
protocol can be replaced by an ideal protocol and no application will
see any significant difference. (We will introduce simulators later.)
Let us consider that we ask some individual, a cryptoanalyst, to
determine whether or not the proposed bit commitment protocol is as
good as an ideal bit commitment. This cryptoanalyst chooses an
application and an adversary circuit $A_r$ for every corruptible
role-circuit $r$ in the protocol. An {\em adversary circuit} $A_r$,
also denoted {\tt Adv}$(r)$, interacts with the application and the
protocol, but is only active when the circuit $r$ is corrupted.  The
circuit $r$ is turned off when it is corrupted.  The adversary circuit
$A_r$ can act in the name of the corrupted circuit $r$ and it has
access to the registers that are normally accessed by $r$.  For
example, the cryptoanalyst will choose an adversary circuit {\tt
Adv}({\tt BC-Alice}) to potentially replace {\tt BC-Alice} and an
adversary circuit {\tt Adv}({\tt BC-Bob}) to potentially replace {\tt
BC-Bob}. With the help of a simple mechanism that we will describe
later, the application can decide at run time which of these two
circuits will be corrupted and replaced, if any.

To have a uniform viewpoint on all adversary circuits, we use the rule
that every adversary circuit $A_r$ is associated with a role-circuit
$r$ in the protocol.  For example, in QKD, we define Eve =
Adv(\mbox{Auxiliary}) where \mbox{Auxiliary} is an
extra circuit in QKD (not owned by Alice or Bob). Note that, in some
QKD protocols, this extra circuit could have an active role in the
(non corrupted) protocol, for example, to create and share EPR pairs
between Alice and Bob.  The set of adversary circuits is called the
{\em adversary}.

The {\em application} is a set of role-circuits that provides inputs
to the bit commitment protocol and receive outputs from this bit
commitment protocol. The application can also communicate with the
adversary circuits that are active.  The application together with the
adversary is called the {\em environment}.  So, the environment
contains many circuits.  Some of these circuits belong to the
application. Some others of these circuits belong to the adversary.
The environment together with the protocol analyzed is called the {\em
{overall setting}}.  Intuitively, the criteria is that, for every
environment, the distribution of a test bit $Z$ is close to an ideal
distribution (associated with an ideal protocol). So, the
cryptoanalyst must pick special applications that can dynamically
corrupt circuits, interact with the associated adversary circuits and,
finally, output a test bit $Z$.  For example, it can be a coin toss
protocol with additional circuits that corrupt role-circuits (i.e.,
turn them off and activate their associated adversary circuits
individually) in the protocol using a simple method described
later. The output bit Z of this application can be the outcome of the
coin toss.

As we previously mentioned, for a general composable definition, to
make sure that no security aspect is ignored, the cryptoanalyst must
consider all possible applications in a large class of applications
that may call this bit commitment.  However, to be concrete, let us
only test how well the bit commitment protocol does when it is called
by a coin toss protocol.  In this coin-toss protocol, Alice picks a
bit $x$ uniformly at random and commits this bit to Bob. Bob waits for
the acknowledgement, then picks a bit $y$ uniformly at random and
announces this bit to Alice. Alice opens the bit $x$ and outputs the
bit $w = x \oplus y$ (the sum modulo 2).  Bob outputs the bit $w = x
\oplus y$ if Alice opening is accepted and opens $y$ otherwise. The
coin-toss module that calls bit commitment is formally defined in
Appendix~\ref{formal_protocols}.  This coin-toss protocol does not
corrupt any party and does not have any mean to corrupt a party
anyway.  So, the cryptoanalyst must pick a variation on this coin toss
protocol that corrupt a party.

Let $w_A$ and $w_B$ be the respective output of Alice and Bob in the
coin toss.  With a simple mechanism that we will describe later, a
coin toss application can corrupt either Alice or Bob or neither of
them.  A coin toss application can corrupt Bob and set $Z = w_A$ to
test the security against Bob who attempts to find out $x$ before the
opening and create a bias on $w_A$.  Another coin toss application can
corrupt Alice and set $Z = w_B$ to test the security against Alice who
attempts to change the value of $x$ to create a bias on
$w_B$. Intuitively, the criteria associated with these two
applications, in both cases, is that there should be no bias on $Z$.
In this way, we have the two fundamental properties of bit commitment:
security against Alice and security against Bob.

\subsection{Why we want to consider a general class of applications?}
\label{GeneralClassOfApplications} 
The above discussion suggests that an interesting definition for bit
commitment can be obtained even if the cryptoanalyst considers only
coin-toss based applications.  If the coin toss protocol is as good
when the analyzed bit commitment is called as when an ideal bit
commitment is called, then we must have an interesting bit
commitment. Really, we must have some non-trivial form of bit
commitment because otherwise the coins toss would be biased. If the bit
commitment is not binding, a corrupted Alice can create a bias on
Bob's output. If it is not concealing, a corrupted Bob can create a
bias on Alice's output.  So, why do we suggest to consider all
applications?  We propose to consider all applications because, even
if we have a specific application in mind, to prove the security of an
application protocol, it is often useful to have stronger results
about the protocols that are called by this application protocol.  So,
the philosophy behind the composability theory is to use strong
security definitions that consider all applications so that at each
level the job is easier. As we will see, the job will be easier
because the security statement is very strong: it says that secure
sub-protocols can be replaced by ideal protocols. This philosophy can
be useful even if at every level in the tree we have a specific
application in mind, and are not interested in all application
protocols. In other words, this is not just a generalization because
we would like to cover all possible application protocols.  Moreover,
a proof that consider all applications is not necessarily more
complicated because it can be easier to avoid technical details that
are associated with a specific application.

A related point is that we must not assume that a few applications are
enough to capture all the security issues associated with a given
task. For example, it is not because the two coin-toss applications
that are discussed above cannot distinguish between the analyzed bit
commitment protocol and an ideal bit commitment that the same is true
for all applications.  The two coin toss applications are not
necessarily representative of all possible applications that might
call the analyzed bit commitment protocol.  Composability is a strong
and natural requirement because it says that no application in a large
class of considered applications can tell the difference between the
analyzed bit commitment protocol and an ideal bit commitment protocol,
not only specific applications such as the coin toss applications.  Of
course, it is useful to consider large but yet restricted class of
environments.  This implies a restriction on the applications as
well. For example, one may consider computationally restricted
environments.  Also, typically one must consider environments in which
the application can only corrupt some fraction of the circuits.  It is
also useful to consider other types of restrictions on the
environments.

\subsection{Why the applications output just a single test bit?}
\label{SingleTestBit}
Another subtle issue is the fact that our proposed criteria depends
only on the probability distribution of a single test bit $Z$. Our
result easily generalizes to the case where a test string $\bf Z$ is
used instead of a single bit as long as the measure $d(P_{\bf Z},
P^{ideal}_{\bf Z}) \in [0, 1]$ that is used to compare the actual
distribution with an ideal distribution respects the triangle
inequality. However, this generalization is problematic because, to be
useful, this generalization will have to be applied systematically for
all protocols, and it is more convenient to consider a single bit.
Fortunately, we see from the coin toss applications that it seems that
a single bit approach does already a very good job.  Moreover, even if
we were to accept the principle that we should compare the
distribution of two strings, in the computational scenario, we will
naturally be back to the one bit case.  Indeed, it is typical in the
computational scenario to compare two string distributions $P_{\bf Z}$
and $P^{ideal}_{\bf Z}$ by considering a polynomial time machine that
receives one of these two strings as an input and returns a single
test bit $Z$. If for all these polynomial time machines, the
distribution of the test bit $Z$ is the same whether the string $\bf
Z$ comes from the ideal case or the analyzed case, we say that the two
cases are computationally indistinguishable. So, if we consider this
polynomial machine as a part of an application, this approach brings
us back to the one bit case. In a non computational scenario, the
situation is similar: the single bit case can be used as a natural way
to implement the string case, that is, to compare the distributions
$P_{\bf Z}$ and $P^{ideal}_{\bf Z}$.  More precisely, for every string
$X \in \{0,1\}^m$, we can consider the application that computes the
parity bit $X \otimes \bf Z$ of the random string $\bf Z$, and require
that the probability distribution $P_{X \otimes \bf Z}$ behaves as in
the ideal case.  If all the single bit distributions $P_{X \otimes \bf
Z}$ behave as in the ideal case, so does the distribution $P_{\bf
Z}$. More precisely, it can be shown that if, for all $X \in
\{0,1\}^m$, $|P_{X \otimes \bf Z}(0) - P^{ideal}_{X \otimes \bf Z}(0)|
\leq \epsilon$, then $\|P_{\bf Z} - P^{ideal}_{\bf Z}\|_2 \leq 2
\epsilon$, where $\| \|_2$ is the $L_2$ norm.

\subsection{The model: an example} \label{running_example}
We have seen two coin toss applications based on the coin toss
protocol.  Here we reconsider these applications to illustrate in more
details the essential of our model and the terminology used.  The
analyzed protocol is again the bit commitment protocol with two
circuits {\tt BC-Alice} and {\tt BC-Bob} for the bit commitment module
and three circuits {\tt SOT-Alice}, {\tt SOT-Bob} and {\tt
SOT-Charlie} for the string-OT protocol.  The application uses two
circuits {\tt CT-Alice} and {\tt CT-Bob} for the coin toss protocol.
The adversary contains four adversary circuits {\tt Adv}({\tt
BC-Alice}), {\tt Adv}({\tt BC-Bob}), {\tt Adv}({\tt SOT-Alice}) and
{\tt Adv}({\tt SOT-Bob}) that are associated with the corruptible
circuits in the protocol.  We recall that {\tt SOT-Charlie} is not
corruptible.  The application has access to the test bit $Z$ and to
four corruption registers: $C$({\tt BC-Alice}), $C$({\tt SOT-Alice}),
$C$({\tt BC-Bob}) and $C$({\tt SOT-Bob}). In our example, only {\tt
BC-Alice} will be corrupted.  Therefore, only the corruption register
$C$({\tt BC-Alice}) and the adversary circuit {\tt Adv}({\tt
BC-Alice}) are needed, but to illustrate the general situation we
included all four corruption registers and all four adversary
circuits. With the help of the corruption registers, the application
decides when to corrupt a circuit in the protocol and which one will
be corrupted.  We assume that every individual gate in each of these
circuits is conditioned by its associated corruption register.

Because it has full control over the corruption registers, the
application can turn these corruptible circuits on and off, and
replace them by their respective adversary circuit when they are
off. In general, the gates of a corruptible circuit $r$ in the
analyzed protocol are conditioned by a corruption register $C_r$. The
gates of the associated adversary circuit $A_r$ are also conditioned
by the corruption register $C_r$, but the other way around: when the
circuit $r$ is on, $A_r$ is off and vice-versa. 

Note that the worst application (against any analyzed protocol) is the
trivial application where every role-circuit simply forwards back and
fourth every input/output message between the analyzed protocol and an
adversary circuit such as Eve that is always active.  This means that
when we analyze the security of a protocol and want a security
statement that is independent of the application protocol, it is
enough to only consider this trivial case.  However, to be concrete,
here we made the choice to consider a less trivial application
protocol that includes the cointoss module with {\tt CT-Alice} and
{\tt CT-Bob}, and we will be consistent with this choice.

Note that, in the partial ordering of the gates in the circuit {\tt
CT} + {\tt BC} + {\tt String-OT}, the output $w_A = x \oplus y$ (in
{\tt CT}) does not have to occur after the opening (in {\tt BC}).
This is because Bob announces $y$ before the opening and the
generation of the random bit $x$ does not have to occur after the
opening. Note that, in a given execution, $w_A = x \oplus y$ might
occur after the opening, but this is irrelevant because only the
constraints that are imposed on the adversary by the partial ordering
of the gates that is defined by the circuit {\tt CT} + {\tt BC} + {\tt
String-OT} matters. Here is how the {overall setting} proceeds.  The
output bit $w_A = x \oplus y$ is sent to the environment of {\tt CT}
as in the honest {\tt CT} protocol. If $w_A = 1$, the application
protocol corrupts {\tt BC-Alice}, that is, it turns off {\tt BC-Alice}
and turns on {\tt Adv}({\tt BC-Alice}) by flipping the corruption
register $C$({\tt BC-Alice}).  At this point, the adversary circuit
{\tt Adv}({\tt BC-Alice}) has access to all the registers of the
corrupted circuit {\tt BC-Alice}, i.e., $x$ and $\hat s = s[x]$, and
can execute the opening in the name of Alice. This adversary circuit
computes $x' = x \oplus 1$, picks a uniformly picked random string $t
\in_R \{0,1\}^k$ and announces $(x', t)$ to {\tt BC-Bob} instead of
$(x, s[x])$ as it would normally happen if Alice was not corrupted.
The application sets $Z = w_B$.  It is not hard to see that $\Pr(Z =
0) = ( 1 + 2^{-k}) 1/2$.

\subsection{The composability security definition}
In the previous example, the environment has created a bias of
$2^{-k}/2$ on $w_B$ toward $0$. This bias on $w_B$ toward $0$ is not
in itself very impressive because we had the power to choose the
application that we want in this environment.  In particular, in the
application, we were free to directly set $w_B = 0$.  However, this
bias is interesting because it allows the environment to distinguish
between the (analyzed) bit commitment protocol and an ideal bit
commitment protocol. We will define in details what is an ideal
protocol later. It is essentially a protocol that executes the task
perfectly with the help of a trusted party and perfect channels. The
goal of the cryptoanalyst (in other words, the goal of an attack) is
not to create a bias on $w_B$ or $w_A$ or on any other bit, but to
find a test bit $Z$ that detect any difference between the analyzed
protocol and the ideal protocol.  Even if Alice is corrupted, as long
as Bob is not corrupted, there should be no bias at all in the output
$w_B$ of the cointoss if an ideal bit commitment is used in this coin
toss application. On the other hand, we just saw that there exists a
bias if the analyzed bit commitment is used. Such a distinction
between the ideal case and the analyzed case would not show up if we
used an application in which Bob sets $w_B = 0$ systematically,
because $w_B = 0$ would systematically hold on both the analyzed and
the ideal sides.  This example is also interesting because the set of
corrupted parties depends on random values in the protocol.  In this
example, because the protocol is perfectly secure against Bob, the
cryptoanalyst could have corrupted {\tt Alice} at the beginning --- no
one else is interesting to corrupt.  However, it is not hard to
conceive an example where there is an advantage to decide later who
should be corrupted.

To detect a bias toward $0$ on $w_B$, we saw that this application
corrupts Alice (i.e. gives her some strategy to create a bias on $\hat
x$ toward $\hat x = y$).  For this application, the criteria that
tolerates no bias is $\Pr(Z = 0) = 1/2$.  (In practice, $|\Pr(Z = 0) -
1/2| \leq \epsilon$, for some small $\epsilon > 0$ would be fine.)
However, we recall that we want to consider all possible application
protocols.  The cryptoanalyst could have designed a biased coin toss
application where Alice picks $x$ with distribution $(p_0, p_1) =
(1/4, 3/4)$ and Bob picks $y$ with distribution $(p'_0, p'_1) = (3/4,
1/4)$.  In this case, even though {\tt BC} is almost perfect, the
distribution of $w_B$ significantly depends on the adversary circuit.
Therefore, the criterion on the distribution of $Z$, which is used to
accept or reject the analyzed bit commitment protocol, must depend on
the environment.

The above discussion suggests that a general criterion on the
probability distribution of the bit $Z$ is to require that, for every
environment, the probability $\Pr(Z_{analyzed} = 0) \equiv \Pr(Z = 0
)$ in the analyzed case is close to the probability $\Pr( Z_{ideal} =
0)$ in the ideal case:
\begin{equation} \label{criteria}
|\Pr(Z_{Analyzed}= 0) - \Pr( Z_{Ideal} = 0)| \leq
\epsilon
\end{equation} 
where $\epsilon > 0$ depends on the security parameter. This criterion
depends on the environment because $\Pr( Z_{Ideal} = 0)$ depends on
both the environment and the ideal protocol. However, we admit that,
at this point, this criteria is not entirely defined and may seem too
strong, too weak or simply obscure because we have not yet really
explained how the reference probability $\Pr( Z_{Ideal} = 0)$ can be
computed, that is, we have not yet really defined $Z_{Ideal}$.

\ignore{ Na\"{\i}vely, we would simply consider an ideal protocol that
corresponds to what we expect from the analyzed protocol and, if an
environment can in any reasonable way distinguish the analyzed
protocol from the ideal protocol in accordance with the criteria
(\ref{criteria}), we would like to conclude that the analyzed protocol
is not secure.  It is when we try to carefully formalize this
na\"{\i}ve formulation of the definition of composability that the
analysis gets a little bit more complicated.}
  
An obvious problem in the definition of $Z_{Ideal}$ is that the
internal communication in the ideal protocol is usually completely
different than the internal communication in the analyzed protocol.
To be useful, an ideal protocol should be as simple as possible.  An
ideal protocol can use hidden (private, authenticated and no traffic
analysis) channels, and a useful ideal protocol will typically do so,
whereas the analyzed protocol might not have access to hidden
channels.  We cannot ignore the internal communication because the
adversary part of the environment can access it. If the adversary
waits for some message and this message does not come because it is
not generated by the ideal protocol, the adversary will jam.  If say
$Z$ is initially set to $0$, it will remain set to $0$ whereas, on the
analyzed side, the environment can be designed to swap $Z$ to $1$ with
some probability. In this way, the analyzed protocol will be rejected
because criteria (\ref{criteria}) is not respected.  With this
na\"{\i}ve approach, even good protocols will be rejected.
 
The solution to this difficulty is to consider a game in which the
players are (1) the environment and (2) a simulator that attempts to
generate the information that is not generated by the ideal protocol
but is normally generated in the analyzed protocol and expected by the
environment.  The ideal protocol and the simulator, both together,
constitute an ``extended ideal protocol''. The environment is the same
in both sides.  The environment tries to distinguish the analyzed protocol
from the extended ideal protocol in accordance with the criteria
(\ref{criteria}). If the environment wins, the protocol is not secure.
The simulator tries to compensate for
\begin{itemize}
\item the missing communication that would normally be generated by
the non-corrupted parties in the analyzed protocol 
and
\item the registers that are normally owned by the corrupted parties
in the analyzed protocol
\end{itemize}
so that the extended ideal protocol and the analyzed protocol cannot
be distinguished. If the simulator wins, the protocol is secure.  Note
that the missing interaction that would normally occur between the
environment and non-corrupted parties in the analyzed protocol can be
separated in two categories.  First, there is the interaction between
the environment and the registers that are exchanged between the non
corrupted parties in the protocol.  This first type of information
depends on the fact that the channels are not private or not
authenticated, etc.  Second, there is the interaction that occurs
because an adversary circuit $A_r$ acts on behalf of a corrupted
circuit $r$.  This second type of interaction can occur even if the
channels are perfect, authenticated, etc. The point here is that the
simulator should compensate for these two types of interaction.

An interesting question can be raised. Why not allowing the adversary
circuit to be different on the ideal side? It turns out that this
alternative definition is almost equivalent to the above definition
(See Appendix \ref{not_stronger} for details.)

\subsection{The main idea behind composability} \label{main_idea}
Thus far, with the help of an example, we discussed a general approach
to define the security of a protocol, but we have not explained the
concept of composability.  This situation is natural because,
historically, the notion of simulator and ideal protocol were first
used in security definitions independently of the concept of
composability.  It is much after that the concept of composability was
explained. Here, we use our example to illustrate the basic idea
behind composability, and its relationship with the proposed
definition.

In our example, the {\tt SOT} sub-protocol is an ideal functionality
with a trusted circuit {\tt SOT-Charlie}.  A more complex bit
commitment protocol would be obtained if we still used the module {\tt
BC}, but replaced the {\tt SOT} ideal sub-protocol with a real
sub-protocol {\tt RealSOT}.  Now, suppose that we have successfully
shown that {\tt RealSOT} securely realizes the ideal protocol {\tt
SOT} in accordance with our proposed security definition.  Suppose in
addition that we have successfully shown that the analyzed bit
commitment protocol which call the ideal {\tt SOT} securely realizes
an ideal bit commitment protocol.  It is much easier to show the
security of ${\tt BC}$ when it calls the ideal sub-protocol {\tt SOT}
than if it called the real sub-protocol {\tt RealSOT} instead.  Great,
but how useful are these security results if we want to obtain the
security of the real bit commitment protocol that calls the real
sub-protocol {\tt RealSOT}?  The main difficulty is that the security
criteria says that the real sub-protocol {\tt RealSOT} can be replaced
by the ideal sub-protocol {\tt SOT}, but with an additional simulator.
The simulator is a problem because we have proven the security of {\tt
BC} which calls the ideal {\tt SOT} without simulator. The main idea
behind the composability definition is that, when we proved the
security of bit commitment protocol that calls the ideal {\tt SOT}, we
considered all environments and, therefore, it is possible to consider
that the simulator is a part of the environment.  We see that it is
important for the composability aspect of the security definition that
the simulator can be interpreted as a part of the environment.  This
is the main requirement for composability.  It is also important that
the simulator together with the ideal protocol can be interpreted as a
protocol, called the extended ideal protocol, because the original
environment (without the simulator) expect an interaction with a
protocol.  However, irrespectively of composability, this second
requirement is a minimal requirement for any simulator that is used to
prove the security of a protocol {\tt RealSOT}.  It is not
specifically related to the composability aspect of the security
definition.

\section{Model for Universal Composability} \label{BasicModel}
This Section describes the formal model used to define the
composability of classical and quantum protocols. The motivation for
this model was provided in the previous Sections.  It should be
pointed out that in practice a protocol corresponds to some source and
the protocol in execution is a different concept, a bound protocol.
The source code does not have fixed participants. They are determined
in the binding.  For simplicity, we assume that the deployment of the
source code into a bound protocol is already executed and we analyse
the bound protocol.  This is natural in a circuit based model. To
simplify the terminology, because we never use the concept of a source
code anyway, the bound protocol that is analysed is simply called a
protocol.

\subsection{Protocols}
A quantum gate is a unitary transformation on a set of registers. A
classical gate is a permutation on the set of classical values of the
registers, a special kind of quantum gate.  (The composability theory
works fine if a classical gate is any function, not necessarily a
permutation, but then we lose the fact that a classical gate is a
special kind of quantum gate.)  A {\em classically controlled gate} is
a quantum gate of the form $\sum_x |x\rangle^X\langle x| \otimes
U^Y_x$.  The unitary transformation $U^Y_x$ is the controlled part of
the controlled gate, the classical register $X$ is the control
register and the register $Y$ is the target register. For simplicity,
we assume that the control register is always a single classical
bit. In other words, whether the controlled part is executed or not is
always precomputed into a single classical bit, the control bit. 

A classical register can be accessed by classical gates and by quantum
gates but, in this latter case, only as a control register or to store
the outcome of a quantum measurement (which can be considered as a
kind of quantum gate).  A quantum measurement always use a fresh
classical register initially set to zero to store its outcome: no
other gate can access this register before the quantum
measurement. Therefore, a classical register can receive the random
outcome of a quantum measurement as initial value, but thereafter it
can only be modified deterministically trough classical gates.

A circuit is a set of gates with a partial ordering that depends on
the initial value of its control registers. For every initial value of
the control registers, the partial ordering must respect a constraint
that we now describe.  For every initial value of every control
register, we consider that a controlled gate {\em accesses} its target
registers only if its controlled part is active, that is, distinct
from the identity. The constraint is that, for every value of the
control registers, every pair of gates that access a same classical or
quantum register must be an ordered pair in the partial ordering. To
compute the complexity of a circuit, we only consider controlled gates
where the controlled part is active. Note that such a complexity is a
value that depends on the initial value of the control registers.

A module-role-circuit is a circuit executed by a given participant. A
protocol is the union of many module-role-circuits.  Each participant
executes one or more module-role-circuits in the protocol.  Two
distinct module-role-circuits can share a common register called a
channel register. As we will see, the adversary can also access these
channel registers.  A communication gate of a module-role-circuit $r$
is a gate of $r$ that accesses a channel register of $r$ and an
activation register associated to this gate.  A transmission
corresponds to a communication gate of a sender module-role-circuit
that accesses a channel register and, next, the activation register of
a (thus subsequent) communication gate of a recipient
module-role-circuit that accesses the same channel register. Only a
communication gate of a module-role-circuit $r$ or a
(non-communication) gate of an adversary circuit can access a channel
register. To allow syncronisation there are special gates that are not
executed until after an associated activation register initially sets
to SLEEP is set to WAKE-UP by another gate.  This corresponds to the
standard approach used in distributed computing in which a procedure
sleeps until it is awaken by another procedure at an appropriate time.

In a well-defined protocol, every channel register has a channel-type
and a channel-register-ID that includes the recipient-ID, the
sender-ID and any other information necessary to uniquely identify the
channel registers.  The gates of an adversary circuit can access a
channel register, but it must be in accordance with its channel-type.
Similarly, we assume that the use of communication gates always
respect the constraints imposed by the channel-type of the
channel-register. For example, if the channel-register is classical
and authenticated, the adversary can read it, but cannot change its
value. If the channel register is quantum and authenticated, the
adversary cannot access it. In both cases, the associated activation
registers are also authenticated, which means that only the legitimate
module-role-circuits can set them to WAKE-UP.

Note that, without channel-types, a protocol specifies no constraint
on the relative order of gates that belong in distinct
module-role-circuits: the sender-ID and the recipient-ID in the
channel-register-ID do not have to respectively correspond to the
actual sender and the actual recipient. If the channel registers are
authenticated, an order is usually implicitly imposed by the protocol
with the help of the activation registers. However, no real channel
register is perfectly authenticated. In reality, the order is not
really fixed: it is only that some test fails if this implicit order
is not respected. Therefore, the union of all module-role-circuits in
a real protocol is not really a circuit: two communication gates that
access a given channel register do not have a fixed order.

A corruptible circuit $r$ is a module-role-circuit that contains a special
bit $C_r$ called the corruption register.  Every gate in the corrupted
circuit $r$ first reads the corruption register $C_r$ and do nothing
else if the corruption register is on. Formally, we consider that
every gate $G$ of a corruptible circuit $r$ has the form
\[
G = |0\rangle^{C_r}\!\langle 0| \otimes G^{honest} +
|1\rangle^{C_r}\!\langle 1| \otimes {\bf I}.
\] 
As discussed in Appendix \ref{more_about_adversary}, it is not hard to
consider different, more nuanced, types of corruption beyond fully
corrupted and not corrupted at all.

A valid environment $E$ for a protocol $P$ is a choice of order for
every pair of communication gates of $P$ that share a channel register
(so that the protocol becomes a circuit in itself) and an additional
circuit with its own set of registers $R_E$ that can also access the
channel registers and the corruption registers of $P$ and is such that
$E + P$ is a circuit (i.e., for every possible value of the control
registers, the union of the two partially ordered set of gates is also
a partially ordered set of gates). The environment contains a special
output register $Z$ and an adversary circuit $A(r)$ for every
corruptible circuit $r$ in $P$. The gates of an adversary circuit
$A(r)$ are like the gates of the corruptible circuit $r$ except that
they are active when the corruption register $C_r$ is on and inactive
otherwise.  Formally, every gate $G'$ of an adversary circuit $A_r$
has the form
\[
G' = |0\rangle^{C_r}\!\langle 0 | \otimes {\bf I} +
|1\rangle^{C_r}\!\langle 1| \otimes G^{\prime corrupted}.
\]
An (active) adversary circuit $A(r)$ can access all registers in the
set registers $R_E$ of the environment and all registers of the
corrupted circuit $r$. In the application of the constraints of a
channel-type, the circuit $A(r)$ can communicate in the name of $r$ as
if it was the circuit $r$. Note that an adversary circuit can
eavesdrop a channel without using the identity of corruptible circuit
$r$. For example, Eve can eavesdrop the communication between Alice
and Bob without using the identity of Alice or Bob.  Note also that we
use the rule that every adversary circuit must be of the form $A(r)$
for some corruptible circuit $r \in P$. For example, $Eve =
A(\mbox{Auxiliary})$ where Auxiliary is an extra
circuit in $P = QKD$.

\subsection{Ideal protocols and simulators}
An ideal protocol $I(P)$ for an analyzed protocol $P$ contains one
circuit $I(r)$ for every module-role-circuit $r \in P^{I/O} \subseteq P$.
Formally, $P^{I/O}$ can be any subset of $P$.  However, the ideal
protocol will only be realizable if $P^{I/O}$ contains every circuit
$r \in {\cal P}$ that uses an input or an output channel to
communicate with the environment. Note that $P^{I/O}$ will often
include the Auxiliary in $P$, even if it does not contribute
to the input/output of the protocol.  We will come back to the special
circuit $I(\mbox{Auxiliary})$ later.  The ideal protocol can,
and usually do, contain other circuits beside the circuits $I(r)$, $r
\in {\cal P}^{I/O}$.  It typically contains one non corruptible
circuit, the so called trusted party.

A circuit $I(r) \in I(P)$ with $r \in {\cal P}^{I/O}$ is corrupted by
the corruption register $C_r$ in the same way as was the circuit $r
\in {\cal P}^{I/O}$ and it can use the input or output communication
channels that are normally used by the circuit $r$ as if it was the
circuit $r$.  However, it does not have access to the internal channel
of the circuit $r$. It uses its own internal channels.

Typically, the ideal protocol $I(P)$ associated with an analyzed
protocol $P$ is much simpler than the protocol $P$. It can accomplish
the task in a much simpler way because it can use ideal channels and a
Non-corruptible party.  In particular, the set of internal channels in
the ideal protocol is usually much smaller or less used than in the
analyzed protocol.  Therefore, if we keep the same environment and
replace $P$ by its ideal protocol $I(P)$, after the corruption of a
circuit $r$, the adversary circuit $A(r)$ in the environment will wait
for ever for messages that will never come, and the output bit $Z$
will never be set as it is set when the environment interacts with
$P$.  Therefore, the two cases will easily be distinguished.  So, it
is not reasonable to require that $P$ and $I(P)$ are not
distinguishable in such a direct manner.  The solution to this dilemma
makes use of a {\em simulator} $S(P)$ that contains one simulator
circuit $Sym(r)$ for every corruptible circuit $r \in P$ and one
adversary circuit $A^{ideal}(r)$ for every corruptible circuit $r \in
I(P)$. The simulator can contain other circuits. 

Each simulated circuit $Sym(r)$ in the simulator, as it is the case
with the circuits of a protocol, is associated with a set of
registers.  This is necessary to determine which registers can be
accessed by the environment when a circuit gets corrupted.
Nevertheless, to simplify the design of the simulator, we allow any
circuit of the simulator to access any register of any other circuit
in the simulator. Through the simulation circuits $Sym(r)$, the
simulator can provide the registers and the communication that is
expected by the adversary circuits $A(r)$ in the environment.  Through
the adversary circuits $A^{ideal}(r)$, the simulator can eavesdrop the
internal communication and impersonate the corrupted circuits in the
ideal protocol.  In this way, the simulator can be seen as an
extension of the environment. The purpose of the circuits
$A^{ideal}(r)$ is to allow the simulator to obtain information that it
needs to accomplish the simulation.  The overall purpose of the
simulator $S({\cal P})$ is to extend the ideal protocol $I(P)$ so that
$E + P$ and $E + I(P) + S(P)$ are indistinguishable. We have the
constraint on $I(P) + S(P)$ that $E + I(P) + S(P)$ must be a circuit,
that is, the union of the partially ordered sets of gates for $E$,
$I(P)$ and $S(P)$ must still be a partially ordered set, every two
gates that share a register must be an ordered pair, etc.  This
suggests the following security definition. In this definition, $n$ is
a security parameter used in the protocol, the simulator and the
adversary circuits. For example, $n$ can correspond to the number of
photons sent in the protocol. The simulator and the adversary interact
with the protocol, so they must also use the security parameter $n$.
\begin{definition} \label{not-hiding-definition}
A protocol $P$ $\epsilon_P$-securely realizes an associated ideal
protocol $I(P)$ if, for every ``valid'' environment $E$ for $P$, there
exists a ``valid'' simulator $S(P)$ such that
\begin{equation} \label{ReplaceFormula1}
|\Pr(Z_{(E + P)} = 0) - \Pr(Z_{(E + I(P) + S(P))} = 0 )| \leq
 \epsilon_P(n, E)
\end{equation}
where $Z_{A}$ is the output bit of $E$ in the overall setting $A$
where $A = E + P$ on the analyzed side and $A = E + I(P) + S(P)$ on
the ideal side. 
\end{definition} 
Of course, in many cases such a simulator only exists when we restrict
the power of the environment. For example, it is typical to assume
that an environment can only corrupt a set of circuits $X \in {\cal X}
\subseteq 2^{\cal P}$ where $2^{\cal P}$ is the set of subsets of
${\cal P}$. The set ${\cal X}$ is called the access rule. 

There is something special about the above definition. The value
$\epsilon_P$ is a function of the environment $E$ which makes it
impossible to evaluate because the environment is
unknown. Fortunately, this dependence on the environment $E$ is not
needed in the case of unconditional security, i.e., in this case we
usually achieve $\epsilon_P(n, E) = \epsilon_P(n)$.  In the case of
computational security, there is no way to achieve the above
definition with an $\epsilon_P$ that does not depend on the
environment, at the least on its size. This is a problem because an
important aspect of universal composability is to have a security
statement that hold for a large class of environments where the size
is unknown.  In practice, the only way out of this problem is simply
to assume some upper bound on the size of the environment so that we
can evaluate $\epsilon_P = \epsilon_P(n, |E|)$. However, in the
security analysis of a complex protocol with many sub-protocols, each
with its associated function $\epsilon$, keeping track of all these
functions $\epsilon$ can be cumbersome. So, for convenience, we simply
require that $\epsilon_P(n, E)$ is a negligible function
$\epsilon_P(n, |E|)$:
\begin{definition} \label{negligible}
A function $\epsilon(n, |E|)$ is negligible (against any polynomial
environment $E$) if, for every polynome $p(n)$, for every $f_n \leq
p(n)$, we have that, for every polynome $q(n)$, for $n$ sufficiently
large, $\epsilon(n, f(n)) \leq 1/q(n)$ or, better, for some $\alpha >
0$, $\epsilon(n, f(n)) \leq 2^{-\alpha n}$ for $n$ sufficiently large.
\end{definition}
The definition in the computational scenario becomes:
\begin{definition} \label{hiding-definition}
A protocol $P$ of polynomial size computationally realizes an
associated ideal protocol $I(P)$ of polynomial size (but usually of
constant size) if there exist a negligible function $\epsilon_P(n,
|E|)$ such that, for every valid environment $E$ for $P$, there exists
a valid simulator $S(P)$ of polynomial size such that
\begin{equation} \label{ReplaceFormula2}
|\Pr(Z_{(E + P)} = 0) - \Pr(Z_{(E + I(P) + S(P))} = 0 )| \leq
 \epsilon_P(n, |E|)
\end{equation}
\end{definition}
Note that, even though it is not explicitly stated that the
environment must be of polynomial size, we need only be concerned with
environment of polynomial size because there is no constraint on
$\epsilon_P(n, |E|)$ when $|E|$ is not bounded by a polyn\^ome.

Actually, to obtain a universal composability theorem, very natural
conditions are added to this definition.  So natural that it maybe
pointless to mention them. Consider the case where many copies of $P$
are used inside some complex protocol.  The polynomial upper bound on
the size of the simulator and the function $\epsilon_P(n, |E|)$ should
be the same for all copies of $P$. In other words, the polynomial
upper bound on the size of the simulator and the function
$\epsilon_P(n, |E|)$ should only depend on the protocol definition,
not on where the protocol is used inside an application protocol.
Moreover, we restrict the universal composability theorem to complex
protocols that use a finite number of distinct protocol definitions
(but it is OK that there are polynomially many copies of each of them.)

\subsection{About the design of an ideal protocol}
As we explained, the input/output communication channels that are
normally available to a circuit $r$ will not be used by the ideal
functionality when $r$ is corrupted.  However, as we previously
explained, it is nevertheless natural to design an ideal protocol that
exchanges with the environment (on the analyzed side) or the simulator
(on the ideal side) the information that would normally be exchanged
through these corrupted input/output channels.  This is not entirely
obvious because, formally, the simulator (or the environment) can only
use adversary circuits $A^{ideal}(r)$ that act in the name of
corrupted circuits $I(r)$ when they interact with the ideal protocol,
and thus, because the circuits $I(r)$ are inside the ideal protocol,
the simulator only expects internal communication.  Here is how it
works. A typical approach to construct an ideal protocol is to use a
dummy party $I(r)$ in the ideal protocol for each circuit $r$ that
participates in the input/output functionality of the analyzed
protocol.  A dummy party $I(r)$ just plays the role of a channel
between $A^{ideal}(r)$ and a trusted party ${\tt Charlie} \in I(P)$
that computes alone the functionality of the ideal protocol. Any input
received by a dummy party is sent to the trusted party and this
trusted also use these dummy parties to return an output.  In this
way, when the dummy circuit $I(r)$ is corrupted, the adversary circuit
$A^{ideal}_r$ can exchange input/output communication with the trusted
circuit in the ideal protocol.

Despite the fact that the simulator can only interact with the ideal
protocol in the name of a corrupted party, if it help in the design of
the simulator, a circuit in the ideal protocol can exchange any
information with the simulator (on the ideal side), even if no circuit
in the analyzed protocol is corrupted.  This exchange can be done
through the circuit $Devil = A^{Ideal}(\mbox{Auxiliary})$ in the
simulator, where \mbox{Auxiliary} is an extra circuit in the
protocol.  For example, this extra circuit can be the same
\mbox{Auxiliary} that is used in the definition $Eve =
Adv(\mbox{Auxiliary})$ in QKD. 

Note that $I(\mbox{Auxiliary})$ in $I(r)$ does not have to be
passive.  It can communicate with the trusted circuit in $I(r)$, even
if the circuit $\mbox{Auxiliary}$ is passive in $r$. We recall that
the environment of the ideal protocol {\em on the ideal side} is the
simulator. Because a more powerful simulator makes a security proof
easier, we want an ideal protocol that gives as much power as possible
to the simulator but we must respect the fact that the ideal protocol
must be useful when used inside an analyzed protocol. The circuit
$I(\mbox{Auxiliary})$ must be defined accordingly.  As a part of a
security proof, an analyzed protocol can contain ideal sub-protocols.
In this case, the environment of the ideal protocol is the environment
of the analyzed protocol.  So the environment communicates with these
ideal sub-protocols when they are on the analyzed side.  If this
exchange of information with the ideal protocol is fine in that
scenario, then it is a valid communication.  So, this exchange of
information between the trusted circuit and $I(\mbox{Auxiliary})$
must respect the spirit of the task. For example, an ideal bit
commitment should not unveil the bit that is committed before the
opening, but it is fine that the trusted circuit in the ideal bit
commitment protocol tells the circuit $A^{Ideal}(\mbox{Auxiliary})$
that Alice has decided to open the bit, even if no party is corrupted.

\section{The universal composability theorem}
Any variation on the universal composability theorem is useful when we
consider protocols $P$ that have many sub-protocols.  A sub-protocol
$Q$ of $P$ has the recursive form
\[
Q = M(Q) + \sum_{R \in {\cal C}(Q)} R
\] 
where ${\cal C}(Q)$ is the set of sub-protocols of $Q$ that are called
by its main module $M(Q)$.  If $Q$ is a primitive then there are no
sub-protocol $R$ in the sum and $Q = M(Q)$.  As we explain in Appendix
\ref{directed_acyclic_case}, we can rewrite a protocol where the set
of modules is a directed acyclic graph, not a tree, as a protocol that
is a tree. So, we do not lose generality when we assume that a
protocol is a tree of modules. Let $I(Q)$ be the ideal protocol
associated with $Q$. Let
\[
\widetilde{Q} = M(Q) + \sum_{R \in {\cal C}(Q)} I(R).
\]  
Here is the composability theorem.
\begin{theorem} \label{StandardTheorem}
Computational scenario: If, for every sub-protocol $Q$ of a protocol
$P$, $\widetilde{Q}$ computationally realizes $I(Q)$, then $P$
computationally realizes $I(P)$.  Unconditional scenario: If, for every
sub-protocol $Q$ of a protocol $P$, $\widetilde{Q}$ $\epsilon_{\tilde
Q}$-securely-realizes $I(Q)$, then $P$ $\epsilon_P$-securely-realized
$I(P)$ where $\epsilon_P = \sum_Q \epsilon_{\widetilde Q}$.
\end{theorem} 
Instead of directly proving the theorem, we will describe a proof
technique that should be used if one wants to keep track of the
function $\epsilon$ (see previous section). This will lead us to a
variation on the universally composability theorem which is more
complicated, less practical, but for which the proof is trivial.
Then, it will be easy to explain how the proof must be modified to
consider the unconditional scenario, and similarly for the
computational scenario.

In the proposed proof technique, we assume that, for every
sub-protocol $Q$ (in the tree of sub-protocols) of $P$, we have that
$\widetilde{Q}$ $\epsilon_{\widetilde{Q}}$-securely realizes $I(Q)$
where $\epsilon_{\widetilde{Q}} = \epsilon_{\widetilde{Q}}(n,
E(\widetilde{Q}))$. The proof technique does not explain how to obtain
this hypothesis.  It explains how to use this hypothesis to obtain
that $P$ $\epsilon_P$-securely realizes $I(P)$ for some function
$\epsilon_P = \epsilon_{P}(n, E(P))$.

As explained before, any approach uses a bottom-up traversal of the
tree of sub-protocols of $P$. The natural way to use
(\ref{ReplaceFormula1}) or (\ref{ReplaceFormula2}) is that, at every
node $Q$ in the bottom-up traversal, the protocol $\widetilde{Q}$ is
replaced by $I({Q}) + S(\widetilde{{Q}})$. This is because
(\ref{ReplaceFormula1}) or (\ref{ReplaceFormula2}) essentially says
that we can do this replacement.  The simulator $S(\widetilde{{Q}})$
for a given the environment $E(\widetilde{{Q}})$ is obtained using the
hypothesis that $\widetilde{{Q}}$
$\epsilon_{\widetilde{{Q}}}$-securely realizes $I({Q})$.  For the
first $Q$ in the bottom-up traversal, we have $E(\widetilde{{Q}}) =
E(P)$, the original environment.  However, as we do replacements,
simulators and ideal protocols are added in the environment
$E(\widetilde{{Q}})$ of the visited protocol $\widetilde{{Q}}$.
Nevertheless, for $E$, $P$, $n$ and $Q$ fixed, $E(\widetilde{Q})$ is a
well defined circuit.

Before the first replacement, the entire setting is $E + P$. After the
last replacement, the entire setting is $E + I(P) + \sum_Q
S(\widetilde{Q})$.  Using the triangle inequality for each
replacement, we obtain that the simulator $S(P) = \sum_Q
S(\widetilde{Q})$ is the required simulator such that
\begin{equation} 
|\Pr(Z_{(E + P)} = 0) - \Pr(Z_{(E + I(P) + S(P))} = 0 )| \leq
 \epsilon(n, E)
\end{equation}
where $\epsilon(n, E) = \sum_Q \epsilon_{\widetilde{Q}}(n,
E(\widetilde{Q})) $. For this last sum to make sense, we need that,
for $P$ fixed, $\epsilon_{\widetilde{Q}}(n, E(\widetilde{Q}))$ is a
function of $n$, $E$ and $Q$. This is the case because
$E(\widetilde{Q}))$ is a function of $n$, $E$ and $P$: it contains
ideal protocols and simulators that can be determined given the
security definition of the sub-protocols $\widetilde{R}$ with $R < Q$
in the traversal order.  So, we have proven the following variation on
the universal composability theorem.
\begin{theorem} \label{ImpracticalTheorem}
If, for every sub-protocol $Q$ of a protocol $P$, $\widetilde{Q}$
$\epsilon_{\widetilde Q}$-securely-realizes $I(Q)$, then $P$
$\epsilon_P$-securely-realized $I(P)$ where  
$\epsilon_P(n, E) =
\sum_Q \epsilon_{\widetilde{Q}}(n, E(\widetilde{Q}))$ in which, as
explained above, $\epsilon_{\widetilde{Q}}(n, E(\widetilde{Q}))$ is a
function of $n$, $E$ and $Q$.
\end{theorem} 
A problem with this variation on the universal composability theorem
and in the way to use it as a proof technique is that the description
of the function $\epsilon_{\widetilde{Q}}(n, E(\widetilde{Q}))$
requires that one keeps track of all the simulators and ideal
protocols in $E(\widetilde{Q})$.  If researchers were to use this
theorem as a proof technique, an intermediary security result for a
sub-protocol $\widetilde{Q}$ would have to include a complete
description of the simulator $S(\widetilde{Q})$ used, and the final
security result would be expressed in terms of a summation $\sum_Q
\epsilon_{\widetilde{Q}}(n, E(\widetilde{Q}))$ which cannot be
simplified because each $\epsilon_{\widetilde Q}$ can depend on
different aspects of the environment $E$.  Fortunately, this issue
disappears in the unconditional scenario case because $\epsilon$ is
independent of the environment and theorem \ref{ImpracticalTheorem}
gives us theorem \ref{StandardTheorem}.  In the computational
scenario, if we want to know $\epsilon$, which at some point one will
want to know, the only way out of this issue is to restrict $\epsilon$
to be a function of the size (or other simple aspects) of $E$ and then
make concrete assumptions so that the different
$\epsilon_{\widetilde{Q}}$ can be computed.

If we just want to know that $\epsilon$ is negligible as in definition
\ref{hiding-definition}, the way to take care of this issue in the
computational scenario is subtler.  We need to show that
$\epsilon_P(n, E) = \sum_Q \epsilon_{\widetilde{Q}}(n,
E(\widetilde{Q}))$ is negligible given that each
$\epsilon_{\widetilde{Q}}(n, E(\widetilde{Q}))$ is negligible.  We
must guarantee that $|E(\widetilde{Q})|$ is of polynomial size because
otherwise the fact that each $\epsilon_{\widetilde{Q}}(n,
E(\widetilde{Q}))$ is negligible cannot be used.  At this point, we
simply use the fact that there are only a constant number of distinct
protocol definitions $\widetilde{Q}$ in the protocol $P$, and each has
a simulator with a polynomial upper bound on its size that depends
only on the definition of $\widetilde{Q}$. We obtain that there exists
a single polynome $p(n)$ that bounds the size of each of the
polynomially many simulators in $E(\widetilde{Q})$.  Similarly, we
must have a single lower bound for all the $\alpha$ in the statement
that all $\epsilon_{\widetilde{Q}}$ are negligible. Again, we use the
fact that that there are only a constant number of distinct protocol
definitions $\widetilde{Q}$ in the protocol $P$ and so only a constant
number of distinct $\epsilon_{\widetilde{Q}}$.

We thank Claude Cr\'epeau and Debbie Leung for useful
discussions. This work has been supported in part by the National
Science Foundation under Grant No. EIA-0086038.


\appendix

\section{History of security definitions for quantum key distribution}
\label{history}
For a long time, researchers only considered attacks that interact
with a single photon or few photons at a time, the individual attacks
(for examples see
\cite{BennettBrassard84,BennettBessetteBrassardSalvailSmolin90,%
MayersSalvail94,FuchsGisinGriffithsNiuPeres97,BechmannGisin99,Lutkenhaus00}).
A special interesting case are the collective attacks where the
interaction is with a single photon at a time with a different probe
for each photon, but there is a final measurement on all the probes
together at the end \cite{BihamMor96}.  The security notion for
quantum key distribution was simply that, for all (individual or
collective) attacks that have an error rate below some finite
threshold, Eve's information must be below some associated finite
threshold.  A private key could then be extracted with standard
privacy amplification techniques
\cite{BennettBrassardCrepeauMaurer95}.  These privacy amplification
techniques being taken for granted, researchers considered the
disturbance/information trade-off as the basic ingredient in the
security analysis of quantum key distribution protocols
\cite{FuchsPeres96}.

This works fine for individual attacks. For non individual attacks,
the formal notion of error rate or disturbance is tricky. For example,
with probability 1/100 Eve could intercept all photons and otherwise
do nothing. The error rate after this attack is 1/100, but yet no
privacy amplification techniques can succesfully extract a private
key. Because of this difficulty and the possible entanglement between
photons created by a non individual attack, the security of quantum
key distribution against all attacks was difficult to prove.

The first security result against all attacks was obtained in
\cite{Yao95} for the quantum oblivious transfer protocol of
\cite{CrepeauKilian88}.  Later, this result was adapted in
\cite{Mayers96} for the quantum key distribution protocol of
\cite{BennettBrassard84}.  Even at the time, without any reference to
quantum error correction, the basic result was that, if the number of
phase flips (i.e., bit flips in the complementary basis) in the raw
key is small, then privacy amplification works. A similar property is
needed in the proofs that are based on Quantum Error Correction (for
example, see \cite{ShorPreskill00}). The main point is that we could
use the number of errors in the raw key, including phase flip errors,
as a measure of Eve's information before privacy amplification. A
small number of phase flip errors was called the small weight property
\cite{Yao95}, and later the small sphere property \cite{Mayers96}.

Note that it is not enough that the expected phase flip error rate is
small, below some fixed small threshold. Privacy amplification only
works when the actual number of phase flip errors, not just its
expected value, is below some small threshold.  For example, in the
above simple attack, the expected phase flip error is 1/100, but
privacy amplification does not work irrespectively of whether or not
1/100 is below any threshold. The solution is to define Eve's success
as the event where the number of bit flips on tested photons is small
and the number of phase flips on the other photons is large. This
leads to a new notion of security: it is not possible for Eve to both
pass the test and have information at the same time. So, the criteria
is of the form $Pr(Pass \wedge Info > \mu) \leq \epsilon$ where $Info$
is some variable that determines how much information Eve has.

Equivalently, the above criteria can be written $Pr(Pass) \times \Pr(
Info > \mu | Pass) \leq \epsilon$, for some $\epsilon, \mu > 0$. Note
that we can set $Info = 0$ when the test fails because no key is
generated. So, in a way, the key point is that we must consider
separately the case where the test passes and the case where the test
fails, and average over the two cases.  Note that the case where Eve
fails the test corresponds to a key of lenght zero. This suggests a
generalization of the above notion: the key point is to consider each
possible lenght of the key separately, and average over all possible
length. This is in accord with the old principle that in key
distribution it is acceptable that Eve's learn the length of the
key. (Otherwise, it is like adding the extra condition that no traffic
analysis is possible, but this might not be achievable with
unconditional security even in the quantum world.) If we apply this
principle to mutual information, the privacy condition becomes $I(K_A,
K_B; V_E | M) \leq \epsilon$ where $K_A, K_B$ are the respective keys
of Alice and Bob, $V_E$ is Eve's view and $M$ is the lenght of the
key. We recall that $I(K_A, K_B; V_E | M) = \sum_m \Pr(M = m) \times
I(K_A, K_B; V_E | M = m)$.

At the same time that the general principle of conditioning over the
length of the key was proposed, it was realised that a small value for
$I(K_A, K_B; V_E | M)$ is not sufficient for privacy (for example, see
\cite{Mayers00,InamoriLutkenhausMayers01}.)  For example, in
principle, Eve could have an attack in which she forces the key to be
a string of zeros. She would not have to receive any data, i.e., the
conditioned mutual information $I(K_A, K_B; V_E | M)$ would vanish,
and yet Eve would know the key.  If this key is used in a
one-time-pad, there is no privacy at all.  So, an extra condition was
added for privacy called the uniformity: the distributions of both
$K_A$ and $K_B$ must be near uniform (conditioned by $M$ as we
explained above). For example, the criteria for the uniformity of the
distribution $P_{K_A}$ of $K_A$ given $M = m$ can be the variational
distance $\|P_{K_A} - 2^{-m}\|_1 = \sum_{k \in \{0,1\}^m } |P_{K_A}(k)
- 2^{-m}|$. In this case, after we condition over $M$, the uniformity
condition becomes $\sum_{m=0}^{\infty} \Pr(M = m) \times \|P_{K_A} -
2^{-m}\|_1 \leq \mu$.

Note that, if we measure non uniformity with the formula $m - H(K_A |
M = m)$, where $H(K_A | M = m)$ is the entropy of $K_A$ given $M = m$,
it is possible to combine small mutual information and uniformity into
a single privacy condition
\cite{Mayers00,InamoriLutkenhausMayers01}: $\sum_m \Pr(M = m) (m
- H(K_A | V_E, M = m)) \leq \eta = \epsilon + \mu$ where $H(K_A | V_E,
M = m)$ is the entropy of $K_A$ conditioned by the random variable
$V_E$ given the event $M = m$.  Of course, it is always understood
that we must also have the correctness condition: $\Pr(K_A \neq K_B)
\leq \beta$. In this case, conditioning over $M$ makes no difference
because $\sum_m \Pr(M = m) \Pr(K_A \neq K_B | M = m) = \Pr(K_A \neq
K_B)$.

\section{The formal protocols} \label{formal_protocols}
In this Section, we formally describe the modules used in this
document.  
\debprotocol{(Ideal) SOT}

\step{\tt SOT-Bob}: For $i = 0,1$ do \{{\tt receive input} ($s[i] \in
\{0,1\}^k$) from {\tt App}; {\tt send} $s[i]$ {\tt to} {\tt
SOT-Charlie}; \};

\step{\tt SOT-Charlie}: For $i = 0, 1$ do \{ {\tt receive} $s[i]$; \};

\step{\tt SOT-Alice}: {\tt receive input} $c \in \{0,1\}$ from {\tt
App}; {\tt send} $c$ {\tt to} {\tt SOT-Charlie}; 

\step{\tt SOT-Charlie}: {\tt receive} $c$; {\tt send} $OK =$ ``ok'' to
{\tt SOT-Bob};

\step{\tt SOT-Bob}: {\tt receive} $OK$; {\tt send output} $OK$ to {\tt
App} ;

\step{\tt SOT-Charlie}: {\tt send} $\hat{s} {\rm :=} s[c]$ to {\tt
SOT-Alice}; 

\step SOT-Alice: {\tt receive} $\hat{\bf s}$; {\tt send output}
$\hat{s}$ to {\tt App};

\finprotocol 

In the ideal ${2 \choose 1}$-String-OT module, the channels are all
hidden.  The bit commitment (BC) protocol calls the
${2 \choose 1}$-String-OT protocol.  Here is the BC module.

\debprotocol{BC on top of SOT}

\nostep \begin{center} Commit Phase \end{center} 
\step{\tt BC-Bob}: For $i = 0, 1$ do \{ {\tt picks} $s[i] \in_R
\{0,1\}^k$; {\tt send input} $s[i]$ to {\tt SOT-Bob}; \}

\step{\tt BC-Alice}: {\tt receive input} $x \in \{0,1\}$; {\tt send
input} $x$ to {\tt SOT-Alice}\\ {\tt receive output} $\hat s$ from
{\tt SOT-Alice};

\step{\tt BC-Bob}: {\tt receive output} $OK$ from {\tt SOT-Bob}; 
{\tt send output} $OK$ to {\tt App};

\nostep\begin{center} Opening \end{center}

\step\label{opening}{\tt BC-Alice}: {\tt receive input} $FOO \in
\{0,1\}$ from {\tt App}; {\tt send} ($x$, $\hat s$) to {\tt
BC-Bob};

\step{\tt BC-Bob}: {\tt receive} ($x$, $\hat s$) from {\tt BC-Alice};
If ($s[x] = \hat s$) do \{ $\{ \hat x := x\}$ \} else \{ $\hat x :=
\perp$ \}; {\tt send output} $\hat x$ to {\tt App};

\finprotocol 

In the BC module, all channels are hidden except the unique {\tt
send/receive} channel which is used by {\tt BC-Alice} to send the
opening information ($x$, $\hat s$) to {\tt BC-Bob}.  This is a public
but authenticated channel.  Here is the CT protocol.

\debprotocol{CT on top of BC} \step {\tt CT-Alice}: {\tt picks} $x
\in_R \{0,1\}$; {\tt send input} $x$ to {\tt BC-Alice};

\step{\tt CT-Bob}: {\tt receive output} $OK$ from {\tt BC-Bob}; {\tt
picks} $y \in_R \{0,1\}$; {\tt send} $y$ to {\tt CT-Alice};

\step{\tt CT-Alice}: {\tt receive} $y$ from {\tt CT-Bob} ; {\tt send
output} $x \oplus y$ to {\tt App}; {\tt send input} ``open''
to {\tt BC-Bob};

\step{\tt CT-Bob}: {\tt receive output} $\hat x$ from {\tt BC-Bob}; 
{\tt send output} $\hat x \oplus y$ to {\tt App};

\finprotocol

In the CT module, all channels are hidden except the two public but
authenticated {\tt send/receive} channels which are used by {\tt
BC-Bob} and {\tt BC-Alice} to send $x_B$ and $x_A$, respectively.

\section{Equivalence with an alternative definition}
\label{not_stronger}
It may seem that only the application should be the same on both
sides, not the adversary.  After all, we saw that the only purpose of
the ideal side is to determine the probability $\Pr(Z_{Ideal} = 0)$
that is acceptable in (\ref{criteria}).  It should not matter that the
adversary is different on the ideal side.  Consider the alternative
definition where, on the ideal side of the game, $\Pr(Z_{Ideal} = 0)$
can be computed with a different adversary than on the analyzed
side. This alternative definition reads as follows. The protocol is
secure if, for all applications $App$ and adversary $Adv$ for $P$,
there exists an ideal adversary $Adv'$ such that $Z_{Analysed}$ and
$Z_{Ideal}$ cannot be distinguished, where $Analysed = App + Adv + P$
and $Ideal = App + Adv' + I(P)$.  In the original security definition
(without the extra existential quantifier on the adversary), we had
$Ideal = App + Adv + S(P) + I(P)$.  In the alternative definition, the
simulator is part of the ideal adversary: two consecutive existential
quantifiers can be replaced by a single one. This alternative
definition is fine because, if the ideal protocol $I(P)$ is really
ideal, no matter what is the ideal adversary $Adv'$ on the ideal side,
the probability $\Pr(Z_{Ideal} = 0)$ should still be a good reference
value. However, in this document, we consider that the adversary is
the same on both sides.  Certainly, the fact that our security
definition avoids this extra existential quantifier can only make it
stronger (more restrictive on the protocols) because it simply means
that we have less freedom for the acceptable value of $\Pr(Z_{Ideal} =
0)$ in (\ref{criteria}), that is, less freedom to accept the protocol.
In the following, we essentially prove that the fact that our security
definition avoids an extra quantifier on the adversary on the ideal
side does not make it strictly stronger, and so these two definitions
are essentially equivalent. 

We say ``essentially'' because there is a subtle issue. The proof that
the alternative definition is equivalent only works if, in this
alternative definition, ``all environments'' means ``all environments
in which the application $App$ is allowed to interact with registers
that are exchanged between non corrupted circuits in the protocol in
accordance with their security type''. In the original definition
(without the extra existential quantifier), these two classes of
environments have the same power, but we do not know this as a fact in
the alternative definition with the existential quantifier on the
adversary on the ideal side.  Therefore, it seems that the alternative
definition (with the existential quantifier on the adversary) might
actually be weaker (less restrictive on the protocols) if we do not
allow the application to interact with the registers that are
exchanged in non secure channels between the non-corrupted parties in
the protocol. Because, by definition, the application does not include
the adversary circuits, this is a reasonable constraint on the
application (even if the channels are public!.)

\ignore{
We explained that an extra quantifier on the adversary on the ideal
side can only help to accept more protocols.  We now prove that an
extra quantifier on the adversary on the ideal side does not help to
accept more protocols, that is, every protocol $P$ that is accepted by
the alternative definition (with the extra existential quantifier) is
also accepted by the original definition.
}

The proof proceeds in three steps. First, we show that, in the
original definition (without the extra existential quantifier), a
protocol is accepted against all environments in which the application
cannot directly interact with channel registers if and only if it is
accepted against all environments in which the application can
directly interact with channel registers.  In other words, in the
original definition, allowing the application to directly interact
with channels registers does not make the environment more powerful.
For the two other steps, when we say ``all environments'' we mean
``all environments in which the application is allowed to directly
interact with channel registers''.  In the second step, we show that
with the original definition, if a protocol $P$ is accepted against a
sub-class of all environments $E'$ that only use a dummy adversary
(defined later), then it is accepted against all environment $E$.
Third, we show that every protocol $P$ that is accepted by the
alternative definition (with the extra existential quantifier) is also
accepted by the original definition against the environments with the
dummy adversary.  There is a subtle point here. This shows the
equivalence of the two definitions, but only if in the alternative
definition (with the extra existential quantifier) ``all
environments'' means ``all environments in which the application is
allowed to directly interact with channel registers''.

\paragraph{First step.} 
The environment can simulate an interaction between the application
and channel registers because the application can send a probe
register to the adversary which can use it as an ancilla to interact
with a channel register and then return the probe register back to the
application.  So, allowing a direct interaction between the
application and channel registers does not make the environment more
powerful.

\paragraph{Second step.} 
The strategy for the second step is that, for every environment $E$
for $P$, we construct an environment $E'$ for $P$ with a dummy
adversary, obtains the simulator $S$ against $E'$ and show that it
works also against $E$. A dummy adversary is an adversary that only
contains communication gates that are executed between the application
and itself or between itself and the protocol. With these gates, a
dummy adversary does nothing except forwarding back and fourth the
communication between the application and the protocol. We construct
the dummy adversary $Dummy$ in $E'$ in the following way.  Let $P$ be
the protocol and $Adv$ the adversary in $E$. We recall that the
protocol $P$ communicates with the adversary $Adv$.  We denote the
communication gates in $Adv$ between $Adv$ and $P$, the $Adv$-$P$
communication gates. An $Adv$-$P$ communication gate is a gate of the
adversary that access a channel-register that is received by a
module-role-circuit of $P$. Let $Dummy$ be the circuit that contains two
$Dummy$-$P$ communication gates for each $Adv$-$P$ communication gate:
one for reception and the other one for forwarding. Let $SAdv$ be the
same as the adversary $Adv$ except that each $Adv$-$S$ communication
gate is replaced by a corresponding an $App$-$Dummy$ communcation
gate. So, $SAdv + Dummy$ is essentially the same as $Adv$, except that
every $Adv$-$S$ communication gate is replaced by three communication
gates, one in $SAdv$ and two in $Dummy$.  The totality of the
application in $E'$ is $App' = App + SAdv$.  The circuit $SAdv$ can be
interpreted as a simulation of $Adv$ inside the application.  We have
constructed $E' = App' + Dummy$.

It is useful to picture the situation in terms of a graph where the
nodes are the gates in $E + P$ with $E = App + Adv$ and the arrows are
the ordered pair of gates in the associated POSET.  In this picture,
we obtain $E' + P$ where $E' = App' + Dummy$ in the following way.
Every arrow between $Adv$ and $P$ is now an arrow between $Dummy$ and
$P$. This is because $Dummy$ is the new advesary that replaces
$Adv$. Moreover, for each such arrow, we replace the arrow between
$Adv$ and $P$ with an arrow between $Adv$ and $Dummy$, reinterpret
$Adv$ as $SAdv$, and add another arrow inside $Dummy$ from the
reception to the fowarding gates. We cannot create a cycle in this
way. So, any valid environment $E$ correspond to a valid environment
$E'$.

To end this second step, we must construct a simulator $S$ against
$E$. By hypothesis, there exists a simulator $S$ against $E'$.  This
simulator that is valid with $E'$ is also valid with $E$ because,
replacing any two consecutive arrows with an intermediary step in the
dummy adversary by the original single arrow cannot create a
cycle. Second, with this same simulator, the computation is the same
with $E$ or $E'$. So, if $S$ succeeds against $E'$, it also succeeds
against $E$. This concludes the second step.

\paragraph{Third step.}
Consider a protocol $P$ that is accepted by the alternative definition
(with the extra quantifier).  We must show that this protocol is also
accepted in the original definition against all environments $E'$.
So, for a given $E' = App' + Dummy$, we must find $S$ such that $Z_{E'
+ P} \approx Z_{E' + I(P) + S}$.  By hypothesis, because $E'$ is a
valid environment against $P$, the protocol $P$ is accepted against
$E'$ in the alternative definition with the extra quantifier.  So, we
have that there exist an adversary $Adv'$ and a simulator $S'$ so that
$Z_{App' + Dummy + P} \approx Z_{App' + Adv' + S' + I(P)}$.  We define
$S$ so that $App' + Adv' + S'$ and $App' + Dummy + S$ are
equivalent. Essentially, $S$ is identical to $Adv' + S'$, but it
communicates with the dummy adversary $Dummy$ instead of directly with
$App'$.  We have that $Z_{E' + P} \approx Z_{App + Dummy + S + I(P)} =
Z_{E' + S + I(P)}$. This concludes the proof.

If we go back to our coin toss environment example, this means that we
can consider that the adversary on the real side which creates the
small bias $2^{-k}$ is in fact a simulated adversary which
communicates with the real protocol through a dummy adversary, but the
simulator on the ideal side is like an adversary that freely interacts
with the ideal protocol and try to play the role of the dummy
adversary, and tries to compensate for differences between the ideal
and the real protocols -- a difficult job because the dummy adversary
is the worst adversary.

\section{More about the adversary status} \label{more_about_adversary}
Usually, the computational basis states of a corruption register
$C(U)$ are the states $|${\tt NOT CORRUPTED}$\rangle$ and $|${\tt
FULLY CORRUPTED}$\rangle$.  However, we will see that we can have
different levels of corruption such as {\tt HONEST BUT CURIOUS}. The
initial state of every register $C(X)$ is $|${\tt NOT
CORRUPTED}$\rangle$.  The gates of every circuit $U \in {\cal P}$ are
always conditionned by the register $C(U)$.  If the register $C(U)$ is
{\tt FULLY CORRUPTED} when $U$ is activated and ready to execute a
gate $G_i$, the gate $G_i$ is ignored.  The next time that it will be
activated, the circuit $U$ will be ready to execute the next gate
$G_{i+1}$ (which will also be ignored if $C(U)$ remains {\tt
CORRUPTED}). The environment has access to the registers of a {\tt
FULLY CORRUPTED} circuit and can communicate in its name.  A circuit
is {\tt HONEST BUT CURIOUS}, if it is not turned off but the
environment can read its classical registers.  A circuit is {\tt FAIR
BUT CURIOUS}, if it can be replaced by an equivalent circuit (no
difference can be seen by the environment unless it reads its
classical registers) and the environment can read its classical
registers. Also, a circuit might not be forced to give its registers
when corrupted. A circuit that is fully corrupted except that its
register are kept private is {\tt CORRUPTED WITHOUT MEMORY}.  In the
following we will restrict ourselves to {\tt NON CORRUPTED} and {\tt
FULLY CORRUPTED} circuits.

\section{The directed acyclic graph case} \label{directed_acyclic_case}
If the graph of sub-protocols is a rooted directed acyclic graph
instead of a tree, an ideal protocol $I(Q)$ for a sub-protocol $Q$ in
the graph can be in the intersection of two protocols $\widetilde{R}$
and $\widetilde{S}$ with $R < S$ in the traversal order.  After we
replace $\widetilde{R}$, which includes $I(Q)$, we have the problem
that $I(Q)$ is not available anymore for $\widetilde{S}$.

Fortunately, this situation will typically occur with a sub-protocol
$Q$ that is designed so that it still provide its functionality $I(Q)$
even after it is called by $R$.  We must use this fact to reorganize
the protocol so that it becomes a tree of sub-protocols instead of a
directed acyclic graph.  For every sub-protocol $Q$ that is called by
many sub-protocols $R_1 < \ldots < R_{n_Q}$, $n_Q > 1$, starting with
those with the longest path from the we do root, we do the
following. Without loss of generality, we assume that the nodes $R_i$
are ordered by the lenght of the longest path from the root to $Q$ in
which they belong. For $i = 1, \ldots n_Q$, we redefine the ideal
protocol of $R_i$ as $I'(R_i) = I(Q) + \sum_{j = 1}^i I(R_j)$, we let
$R_{i + 1}$ call $R_i$ instead of $Q$ and every protocol $T$ that
calls $R_i$ must now call $R_m$ which is the latest $R_j$, with $j
\geq i$, that occurs before $T$ in the traversal order.

The hypothesis of the universal composability for the new tree will
have to be proven. This may seem a difficult task because of the
apparent complexity of the new ideal protocols $I'(R_i)$.  However,
this complexity is only superficial. The protocol $R_{i+1}$ only uses
$I(Q)$ inside $I'(R_i)$ and any other protocol that called $R_i$ but
now calls $R_m$, only uses $I(R_i)$ inside $I'(R_m)$. This suggests
that whenever we have a directed acyclic graph of sub-protocols, there
are good chances that we can reorganize it as a tree and still be able
to use the universal composability theorem.

\end{document}